\documentclass[aps,prb,twocolumn,superscriptaddress,floatfix]{revtex4-1}
\usepackage{amsfonts,bm}
\usepackage{amsmath}
\usepackage{amssymb}
\usepackage{amsthm}
\usepackage{graphicx}
\usepackage{graphics}
\usepackage{subfigure}
\usepackage{color}
\usepackage{bm}
\usepackage{rotating}
\usepackage{comment}
\usepackage{indentfirst}
\usepackage[normalem]{ulem}

\usepackage{slashed}

\usepackage{float}
\usepackage[titletoc,title]{appendix}

\newcommand{\Z}{\mathbb{Z}}

\newcommand{\ket}[1]{|#1\rangle}

\newcommand{\la}{\langle}
\newcommand{\ra}{\rangle}

\graphicspath{{Z2figures/}}

\begin{document}

\title{Discrete spin structures and commuting projector models for 2d fermionic symmetry protected topological phases}
\author{Nicolas Tarantino}
\author{Lukasz Fidkowski}
\affiliation{Department of Physics and Astronomy, Stony Brook University, Stony Brook, NY 11794-3800, USA}

\begin{abstract}
We construct exactly solved commuting projector Hamiltonian lattice models for all known 2+1d fermionic symmetry protected topological phases (SPTs) with on-site unitary symmetry group $G_f = G \times \Z_2^f$, where $G$ is finite and $\Z_2^f$ is the fermion parity symmetry.  In particular, our models transcend the class of group supercohomology models, which realize some, but not all, fermionic SPTs in 2+1d.  A natural ingredient in our construction is a discrete form of the spin structure of the 2d spatial surface $M$ on which our model is defined, namely a `Kasteleyn' orientation of a certain graph associated with the lattice.  As a special case, our construction yields commuting projector models for all $8$ members of the $\Z_8$ classification of 2d fermionic SPTs with $G = \Z_2$.
\end{abstract}

\maketitle

\section{Introduction}

Recently, it has been realized that gapped phases of matter can be distinguished on the basis of symmetry, even when this symmetry is unbroken.  In particular, gapped lattice Hamiltonians which can be continuously connected to a trivial decoupled `atomic insulator' limit can define distinct symmetry protected topological (SPT) phases \cite{Chen1d, Fidkowski1d, TurnerBerg, Chen2d} when a global symmetry is imposed.  The classification of these SPT phases is well understood in the case of free fermions, where it yields the familiar classification of band topological insulators and superconductors \cite{Ludwig, Kitaev_PT}, but is more difficult for interacting systems.  One approach to classifying interacting SPTs, valid for bosonic systems with discrete on-site unitary symmetry, is to gauge the symmetry, resulting in a model whose low energy physics is described by a topological quantum field theory (TQFT), specifically a discrete gauge theory parametrized by a discrete invariant called a group cohomology class \cite{D_W}.  One can then straightforwardly construct exactly soluble commuting projector lattice models that `gauge' into all possible group cohomology classes \cite{LevinGu, Chen2d, Wang2014a}.  For interacting fermionic systems, however, this strategy is complicated by the fact that upon gauging one expects a spin-TQFT, which requires a spin structure for the manifold on which the low energy field theory lives \cite{G_K, Walker}.  How does this spin structure data enter into the discrete lattice fermionic SPT Hamiltonian?

In this paper, we use a discrete lattice analogue of a spin structure, called a `Kasteleyn' orientation, to write down exactly soluble lattice Hamiltonians for all known 2+1d fermionic SPTs with on-site symmetry group $G_f = G \times \Z_2^f$\cite{M_Cheng, Kapustin2015}.  The introduction of a Kasteleyn orientation allows us to put our exactly soluble models on arbitrary genus 2d surfaces.  We initially focus on the case $G= \Z_2$, and write down exactly soluble commuting projector models that realize all $8$ interacting SPT phases in this case \cite{GuLevin, Qi_Z8, Ryu_Z8, Yao_Z8}.  In particular, our odd $\nu$ commuting projector models, with $\nu$ being the $\Z_8$ index, transcend the group supercohomology class of Hamiltonians introduced in [\onlinecite{supercohomology}].  We then extend our results to all 2+1d fermionic SPTs.  Specifically, we construct models that realize the `root' fermionic SPT phases in the language of [\onlinecite{M_Cheng}]; all other fermionic SPT phases can be obtained by stacking a root phase with a group supercohomology model, the latter having a commuting projector representation as shown in [\onlinecite{supercohomology}].

It is worthwhile to compare our work with that of [{\onlinecite{G_K}], who introduce a different discrete version of a spin structure to study the group supercohomology models.  Namely, [{\onlinecite{G_K}] generalizes the group supercohomology construction of [\onlinecite{supercohomology}] from a fixed triangulation of flat 2+1d space-time to a arbitrary triangulations and topologies\footnote{The topology has to be that of a spin manifold.}, at the expense of introducing additional discrete data encoding a spin structure.  In a Hamiltonian formulation with a trivalent lattice $L$ on a 2d spatial surface $M$, this data amounts to a choice of a subset $E$ of links with the property that the boundary of each plaquette contains an odd number of links of $E$.  This is not the same as a Kasteleyn orientation, even though both encode a choice of spin structure of $M$.  Indeed, whereas the vertices of $L$ correspond to physical fermions, the beyond group supercohomology Hamiltonians we construct are most naturally formulated in terms of a graph $\Lambda$ whose vertices represent Majorana fermions.  It is this graph $\Lambda$ that carries the Kasteleyn orientation.  We will clarify the relation between $L$ and $\Lambda$ below.

The existence of commuting projector Hamiltonians for SPTs is important for several reasons.  First, it has been proposed that many body localized (MBL) \cite{Basko, Pal} phases exhibiting SPT order in generic finite energy density eigenstates \cite{1d_MBL_SPT_1, 1d_MBL_SPT_2, Chandran1, Bauer1} can exist only when the SPT in question has a commuting projector representation \cite{Potter_Vishwanath}.  An odd $\nu$ $\Z_2$ fermionic SPT is interesting in this regard from a quantum computing perspective because an extrinsic $\Z_2$ symmetry flux binds a Majorana zero mode in such an SPT.  Second, a commuting projector Hamiltonian implies an efficient tensor network state (TNS) representation of the ground state and the existence of a gapped parent Hamiltonian for this TNS, allowing such SPT ground states to be targeted by numerical algorithms.  We note that the free fermion realizations of the 2+1d $\Z_2$ fermion SPTs also have TNS representations, but, being composed of decoupled layers of $p\pm ip$ superconductors, the corresponding TNS parent Hamiltonians are necessarily gapless \cite{Read}.

Although the existence of commuting projector Hamiltonians for the odd $\nu$ $\Z_2$ fermionic SPTs may be unexpected \cite{Potter_Vishwanath}, there are some reasons to suggest that it should not be surprising.  For one thing, gauging the fermion parity symmetry in such SPTs results in a bosonic model with toric code topological order \cite{Kitaev2006}, with the $\Z_2$ global symmetry exchanging the $e$ and $m$ excitations.  It is known that a commuting projector model with an onsite $\Z_2$ symmetry acting this way exists \cite{Heinrich}.  Second, gauging the global $\Z_2$ symmetry instead of the fermion parity results in a fermionic theory whose topological content consists of quasiparticles $\{1,\sigma,\psi \} \times \{1,f\}$, where $1,\sigma,\psi$ obey the Ising fusion rules and $f$ is the fundamental fermion \footnote{Indeed, this can be seen directly from the free fermion representation of this particular SPT as two decoupled $p\pm ip$ layers with Chern numbers $\nu, -\nu$ respectively, since here the global $\Z_2$ symmetry just measures the fermion parity of one layer.  The $1,\sigma,\psi$ quasiparticles must form either the Ising or the $SU(2)_2$ modular theory.}.  Since the chiral central charge of the Ising sector, being an integer multiple of $1/2$, can be screened by an appropriate fermionic BdG band structure, there is no chirality obstruction to a commuting projector model of this fermionic topological order, and indeed such a model appears to have been constructed by K. Walker \cite{Walker}.  The method of [\onlinecite{Walker}] involves starting with a doubled Ising string net model and effectively `ungauging' fermion parity by condensing a bound state of the doubled Ising emergent fermion and a fundamental fermion.  The latter also requires a choice of spin structure, which manifests itself in the phase factors associated to various terms in the Hamiltonian in the prescription of [\onlinecite{Walker}].  It would be interesting to try to relate the $\Z_2$ gauged version of our model to that of [\onlinecite{Walker}].

The rest of our paper proceeds as follows.  In section \ref{sec2} we introduce the degrees of freedom of our model for the case $G=\Z_2$, which consist of both spin $\frac{1}{2}$'s and fundamental fermions.  The model is a version of the decorated domain wall construction\cite{Chen2013b} in which the domain walls between the spins bind Majorana chains\cite{Potter_Vishwanath}.  In section \ref{sec3} we introduce the notion of a Kasteleyn orientation and discuss its relation to the spin structure.  In our decorated domain wall model context, the Kasteleyn orientation will be critical in ensuring that the various domain wall configurations between which the ground state fluctuates all correspond to fermionic states with the same fermion parity.  In section \ref{sec4} we write down the Hamiltonian of our model.  The most technical part of this section is the construction of the term that makes the domain walls fluctuate and correspondingly rearranges the Majorana chains.  In section \ref{sec_comm} we prove that all of the terms in our Hamiltonian commute.  In section \ref{sec_analysis} we show that our Hamiltonian does indeed describe an SPT, in the sense that, upon breaking the global $\Z_2$ symmetry, it can be continuously connected to a trivial tensor product state.  We also analyze its SPT order, showing both that an extrinsic $\Z_2$ symmetry flux binds a Majorana zero mode, and that a fermion parity $\pi$ flux binds a doubly degenerate state on which the global $\Z_2$ symmetry and fermion parity anti-commute (i.e. the global $\Z_2$ symmetry acts as an odd operator near such a $\pi$ flux).  Either of these two properties alone establishes our model as an odd $\nu$ $\Z_2$ fermionic SPT.  We also extend our method to show that arbitrary 2+1d fermionic SPTs have commuting projector representations.  We conclude in section \ref{sec_future} with ideas for future directions.

After the completion of this work, we were made aware of a forthcoming paper by Bhardwaj, Gaiotto, and Kapustin\cite{Kapustin_new} which also in particular constructs lattice models for the ‘root’ fermionic SPTs. Shortly after posting, we were also informed about independent work by Ware, Son, Cheng, Mishmash, Alicea, and Bauer\cite{WareSon2016}, who construct two different un-frustrated Hamiltonians, a generalization of the triangular lattice dimer model and a hexagonal Majorana loop model, both realizing the $\text{Ising} \times (p_x - i p_y)$ phase that can be obtained by gauging the $\Z_2$ global symmetry in our model.  In particular, reference \onlinecite{WareSon2016} also introduces a Kasteleyn orientation in the hexagonal model in order to conserve fermion parity under fluctuations of the Majorana loops. 

\section{Lattice model of 2+1d $\Z_2$ fermionic SPT} \label{sec2}

\begin{figure}[htbp]
\begin{center}
\includegraphics[width=0.36\textwidth]{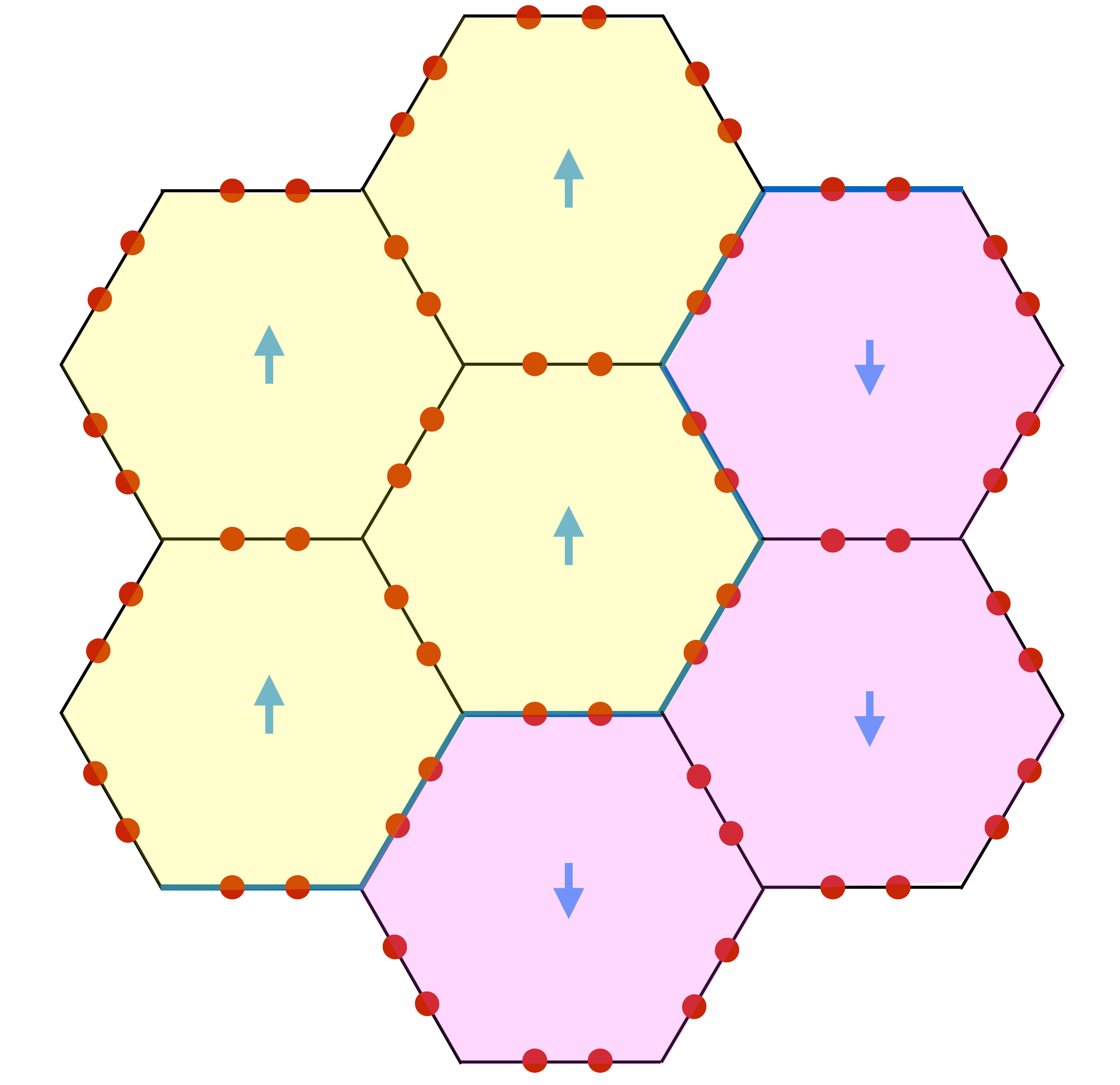}
\caption{Graphical representation of the degrees of freedom in our model.  There is one spinless fermion per link, represented by two Majorana operators, drawn as two red dots.  There is also an Ising spin $\frac{1}{2}$ degree of freedom on each plaquette, represented by a blue arrow.  We will discuss various spin configurations in the $\sigma^z$ basis.  The yellow region represents a spin up domain, the purple region a spin down domain; the two are separated by a domain wall.}
\label{majoranas_without_arrows}
\end{center}
\end{figure}

\begin{figure}[htbp]
\includegraphics[width=0.4\textwidth]{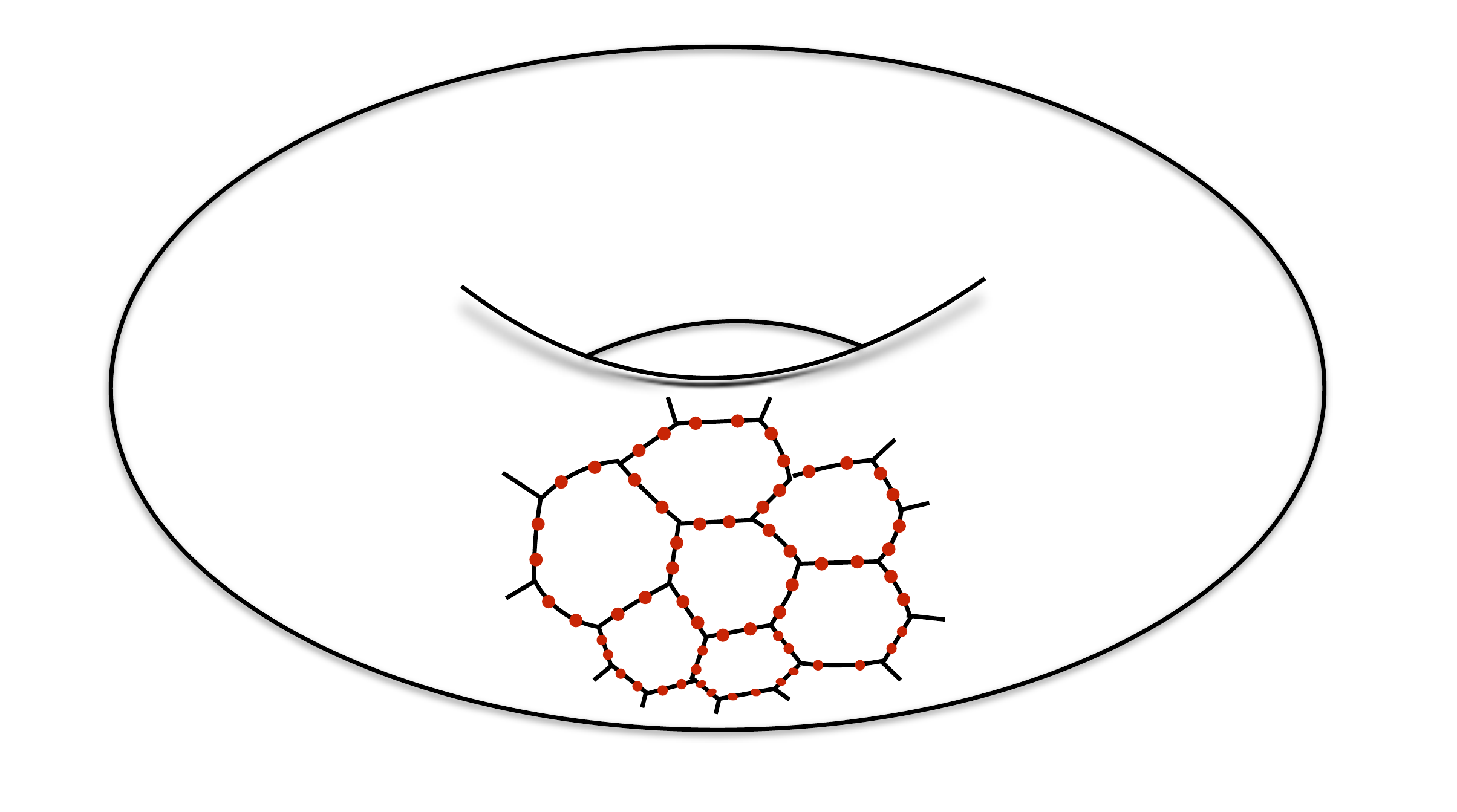}
\centering
\caption{Trivalent lattice on the surface of a torus.  The red dots represent Majorana operators, as discussed in the text.  Although our subsequent discussion is illustrated only on the hexagonal lattice, it applies to general trivalent lattices on arbitrary genus $g$ 2d surfaces $M$.}
\label{new_lattice_on_torus}
\end{figure}

\begin{figure}[htbp]
\begin{center}
\includegraphics[width=0.36\textwidth]{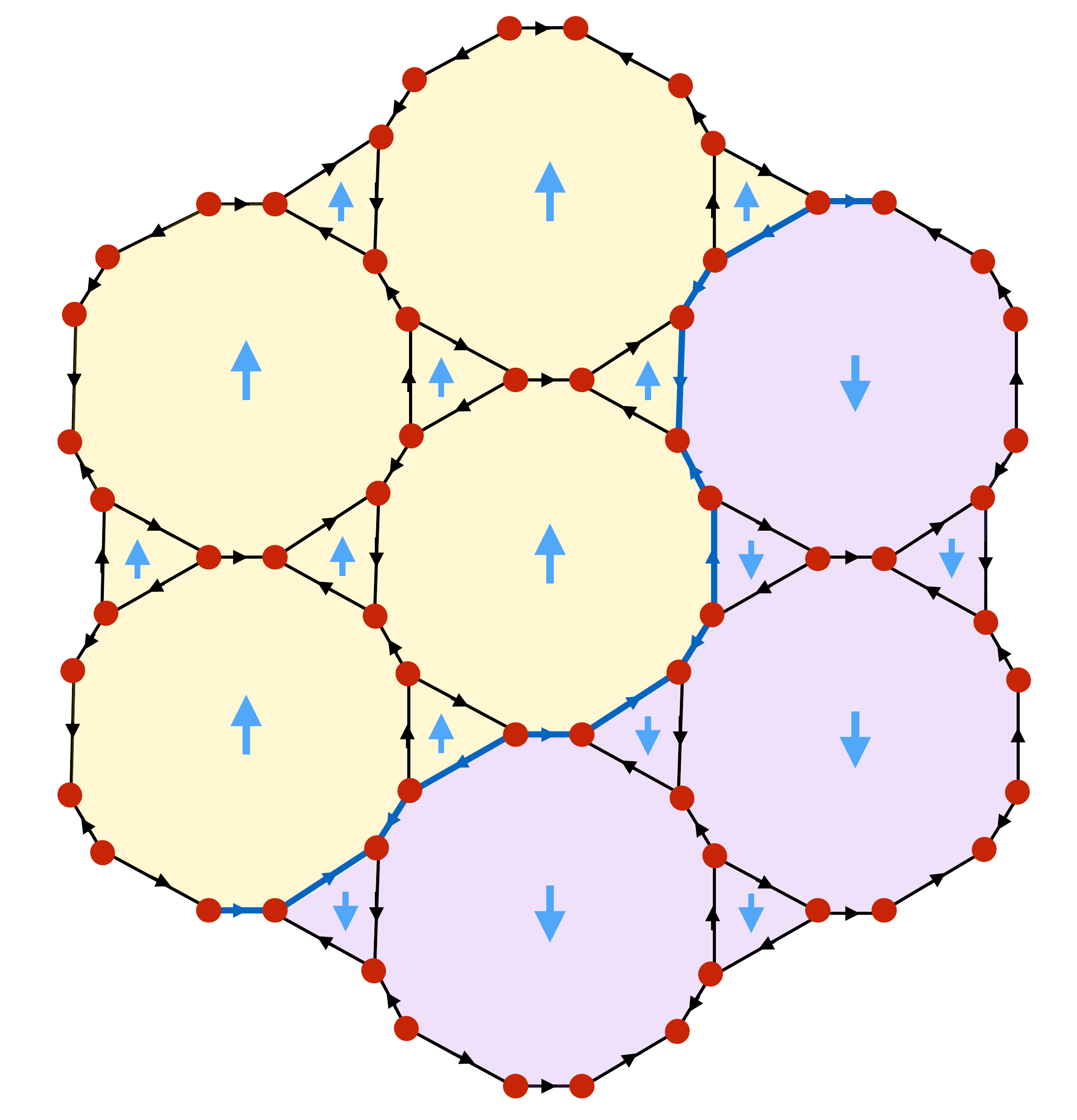}
\caption{Alternative representation of the degrees of freedom in terms of a graph $\Lambda$.  The vertices are the Majorana modes, still represented by the same red dots as before.  Each site of the original lattice $L$ has now been replaced by a small triangular face.  Although the dynamical Ising spin $\frac{1}{2}$ degrees of freedom are located only on the non-triangular faces, which we refer to as plaquettes, we find it useful to extend each spin configuration to a spin configuration over the triangles as well.  This extension is determined uniquely by majority rule: each triangle spin points in the same direction as the majority of its neighbors.  The edges of $\Lambda$ carry a Kasteleyn orientation that preserves the translational symmetry and some of the translational symmetry of $\Lambda$.}
\label{new_full_lattice}
\end{center}
\end{figure}

The degrees of freedom in our model are spin-$\frac{1}{2}$'s located on the plaquettes of a planar trivalent lattice $L$ and spinless fermions located on its links - see figure \ref{majoranas_without_arrows}.  We will often work with a regular hexagonal lattice for simplicity, but our construction works for all planar trivalent lattices.  This is important because we will consider lattices on arbitrary 2d oriented genus $g$ surfaces $M$, as in figure \ref{new_lattice_on_torus}; such lattices cannot be strictly regular hexagonal when $g \neq 1$.  We write $\tau^x_p, \tau^y_p, \tau^z_p$ for the Pauli operators acting on the spin located on plaquette $p$.  The spinless fermion on link $l$ is created and annihilated by operators $c^\dag_l$ and $c_l$ respectively.  However, we will find it more convenient to work with the Majorana combinations $c^\dag_l+c_l$ and $i(c^\dag_l - c_l)$.  In figure \ref{majoranas_without_arrows} these Majorana operators are represented as red dots, with two red dots per link, and the spins are represented by blue arrows on plaquettes.  The global $\Z_2 = \{1,g\}$ symmetry operator is defined to flip the spins:

\begin{equation}
U_g = \prod_p \tau^x_p
\end{equation}
We now introduce an especially convenient representation of our physical system, shown in figure \ref{new_full_lattice}.  This is a planar graph $\Lambda$ based on our original lattice $L$, with each red dot of $L$ corresponding to a separate vertex $v$ of $\Lambda$.  Each trivalent lattice site of $L$ is thus split into three vertices defining a triangular face $t$ in $\Lambda$ - see figure \ref{new_full_lattice}.  Let $t(v)$ denote the triangular face that includes the vertex $v$.  Note that all the other faces of $\Lambda$ correspond one to one with the plaquettes of the original lattice (e.g. the hexagons in the hexagonal lattice correspond to $12$-sided faces in $\Lambda$).  We will thus continue to refer to such faces as plaquettes.  The edges of $\Lambda$ also come in two types: there are edges $\la v w \ra$ connecting different triangles ($t(v)\neq t(w)$), which we call `type I', and the edges within the same triangle ($t(v)=t(w)$), which we call `type II'.  Like the original lattice $L$, the graph $\Lambda$ lives on the 2d surface $M$.

Note that the spin $\frac{1}{2}$ degrees of freedom are defined only on the plaquettes of $\Lambda$, and not at the triangles $t$.  Nevertheless, it will be useful to also define a fictitious spin $\frac{1}{2}$ degree of freedom on each triangle $t$, whose $\tau_t^z$ value is determined according to the majority rule: $\tau_t^z$ is $+1$ or $-1$ depending on whether the majority of the three plaquettes $p$ bordering $t$ have $\tau_p^z=+1$ or $\tau_p^z=-1$.  Thus any spin configuration on the plaquettes extends uniquely to a spin configuration on all of the faces of $\Lambda$, as shown in figure \ref{new_full_lattice}.  Notice that we do not define $\tau_t^{x,y}$ operators, and that $\tau_t^z$ do not correspond to additional dynamical degrees of freedom.  Again, we would like to emphasize that we are only working with hexagonal lattices for definiteness, and everything we have done so far works for arbitrary trivalent lattices.  With this rule, we see that domain walls between different spin configurations always `cut corners', as illustrated in figure \ref{new_full_lattice}, so that not every loop in $\Lambda$ is a valid domain wall.  

A key fact for us will be that there is a one to one correspondence between valid domain wall configurations and dimer coverings of $\Lambda$.  Indeed, given a domain wall configuration, the prescription for extracting the dimer covering is as follows: for a type I edge $\la v w \ra$ (i.e. $t(v) \neq t(w)$), we pair up $v$ and $w$ into a dimer if there is no domain wall along $\la v w \ra$, and for a type II edge $\la v' w' \ra$ (i.e. $t(v')=t(w')$), we pair up $v'$ and $w'$ into a dimer if there is a domain wall along $\la v' w' \ra$.  It is easy to see that this rule gives a valid dimer covering, as illustrated in figure \ref{new_domain_wall}. \footnote{Note that not every dimer covering comes from a loop configuration; although each such covering locally looks like a domain wall boundary, it might not be so globally: e.g. it could be a single loop around a non-trivial cycle on a torus.}

In our construction so far there is an ambiguity as to which Majorana operator - $c^\dag_l+c_l$ or $i(c^\dag_l - c_l)$ - is represented by which of the two dots on each link.  To resolve it, we will introduce an orientation on the edges of $\Lambda$.  Then, for any type $I$ edge $\la v w \ra$ (i.e. with $t(v) \neq t(w)$) oriented from $v$ to $w$, we define Majorana operators $\gamma_v$ and $\gamma_w$ associated with these two vertices by:

\begin{align}\label{cMaj} 
\gamma_v &= c^\dag_l + c_l \\
\gamma_w &= i(c^\dag_l - c_l)
\end{align}
However, for what follows we cannot just choose an arbitrary orientation.  Instead, we require a Kasteleyn orientation, namely one that satisfies the following property: for any face of $\Lambda$ (including the triangular faces), the number of clockwise-oriented edges bounding it must be odd (see figure \ref{new_full_lattice}).  Before explaining why this Kasteleyn property is required, let us discuss some general facts about Kasteleyn orientations.

\section{Kasteleyn orientations and spin structures} \label{sec3}

We now summarize some basic facts about Kasteleyn orientations \cite{Kast}, i.e. orientations of planar graphs for which any face has an odd number of clockwise-oriented edges:

{\bf (1)} A Kasteleyn orientation exists for any planar graph with an even number of vertices, on any genus oriented surface.

{\bf (2)} Given any Kasteleyn orientation, one can obtain another one by flipping the orientations of all edges adjoining any given vertex $v$.  Two Kasteleyn orientations related by a sequence of such moves (with different $v$) are said to be equivalent.

{\bf (3)} There are exactly $2^{2g}$ inequivalent Kasteleyn orientations for a planar graph on a genus $g$ surface\footnote{The graph has to form a sufficiently fine discretization of this surface, to avoid pathological examples; in particular, any cycle on the surface has to be homologous to some cycle on the graph.}.

While the proof of (1) is not completely trivial - see [\onlinecite{Kast}] - (2) and (3) are fairly easy to understand.  Indeed, the difference between any two Kasteleyn orientations - that is, the set of edges where the orientations differ - must define a flat $\Z_2$ gauge field configuration on $\Lambda$, and conversely one can deform any Kasteleyn orientation by any flat $\Z_2$ gauge field.

Fact (3) suggests a connection between Kasteleyn orientations and spin structures, of which there are also $2^{2g}$ on a genus $g$ surface.  A spin structure is just a consistent set of rules for assigning sign factors to fermions moving along framed paths.  In general it is defined as a certain double cover of the `frame bundle' \footnote{A spin structure for a $n$-dimensional manifold $X$ is defined as a principal $\text{Spin}(n)$ bundle $P \rightarrow X$ with a two-fold covering map of bundles $P\rightarrow P_{\text{SO}}$ that restricts to the two-fold covering map $\text{Spin}(n)\rightarrow \text{SO}(n)$ on fibers, where $P_{\text{SO}}$ is the frame bundle of $X$}, but for an oriented $2d$ surface $M$ a more concrete definition exists: a spin structure is given by a non-vanishing vector field on $M$ with only even singularities.  Indeed, such a vector field gives a local system of coordinates with respect to which rotation can be measured, with its even singularities being invisible to fermions.  Because one can modify a spin structure by any flat $\Z_2$ gauge field - now thought of as threading $\pi$ fluxes for fermions through various cycles of $M$ 
- there are also $2^{2g}$ inequivalent spin structures.

Even though both the set of Kasteleyn orientations and the set of spin structures correspond one-to-one with the set of flat $\Z_2$ gauge field configurations, these correspondences are not canonical.  That is, there is no preferred way to choose a Kasteleyn orientation or a spin structure to correspond to the zero $\Z_2$ gauge field configuration \footnote{Formally, both form a torsor over the $\Z_2$-valued cohomology of M, $H^1(M,\Z_2)$.  In fact both correspond to quadratic forms over $H_1(M,\Z_2)$ with bilinear form equal to the intersection pairing.}.  However, importantly for us, given a fixed dimer covering of the graph, the correspondence between Kasteleyn orientations and spin structures {\emph{is}} canonical.  In our graph $\Lambda$ there is a natural dimer covering given by pairing the Majoranas into the original link fermions (eq. \ref{cMaj}), so that a choice of Kasteleyn orientation on $\Lambda$ is really the same thing as a choice of spin structure on $M$.

To elucidate this connection, we will now construct, given a dimer covering and a Kasteleyn orientation of $\Lambda$, the corresponding non-vanishing vector field on $M$ containing only even singularities.  This construction is due to Kuperberg, see [\onlinecite{Kuperberg}] or section 4.3 of [\onlinecite{Kast}].  First, we define the vector field in the vicinity of all vertices to point towards those vertices, then extend over each edge in a manner prescribed by the Kasteleyn orientation, and finally extend over faces, as in figure \ref{new_Kasteleyn}.  The Kasteleyn property ensures that when extending over faces, one encounters only even singularities.  By pairing up the index $1$ singularities around vertices as dictated by the dimer covering one ends up with a non-vanishing vector field with only even singularities, which uniquely defines a spin structure.  Note that the index $1$ singularity around each vertex is consistent with the identification of vertices with Majorana fermion zero modes.

\begin{figure}[htbp]
\begin{center}
\includegraphics[width=0.36\textwidth]{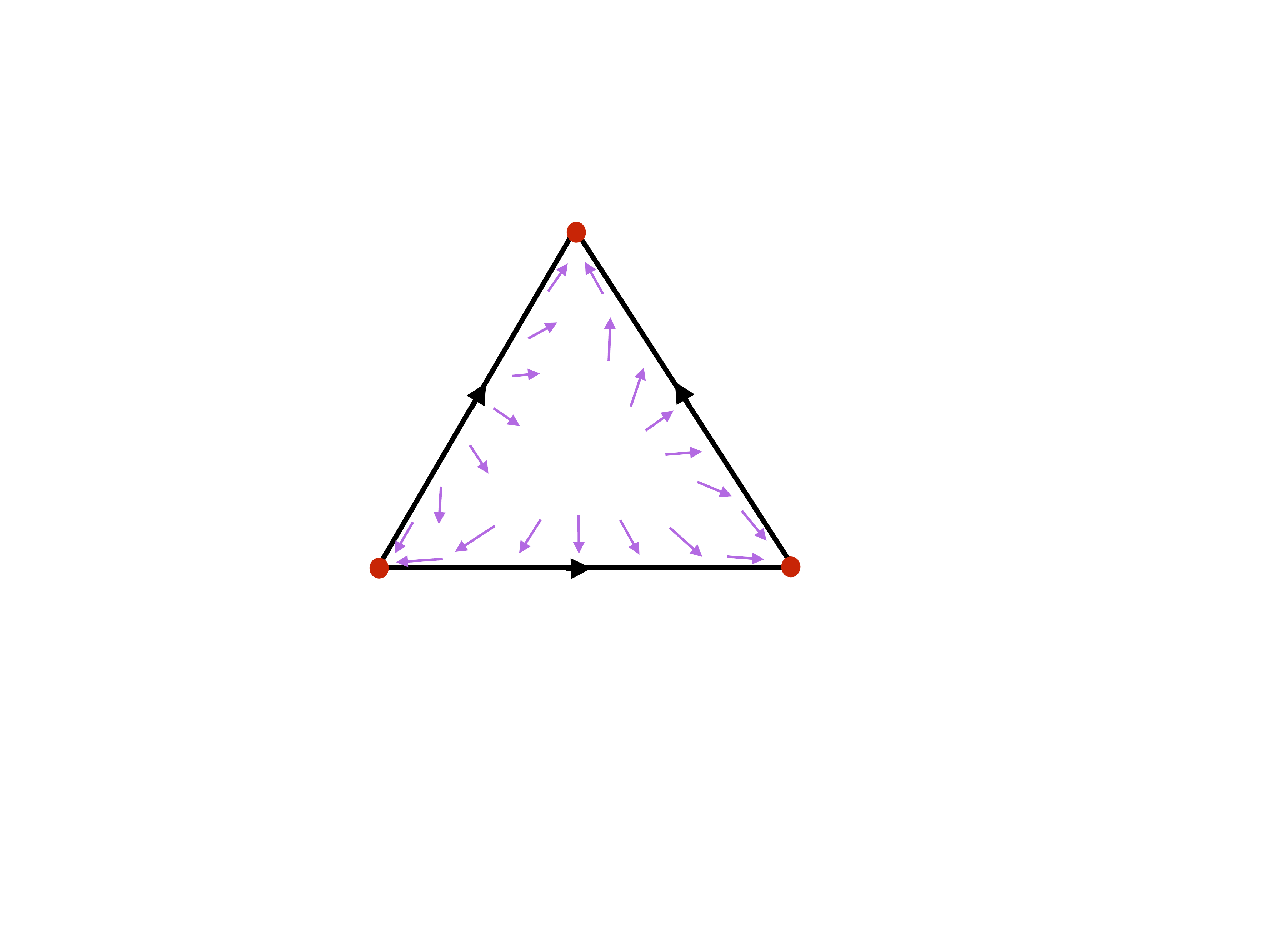}
\caption{Constructing a non-vanishing vector field (purple) with only even singularities, given a Kasteleyn orientation.  The vector field is defined to point towards the red vertices in their vicinity, and is then extended over the edges.  This extension is dictated by the Kasteleyn orientation of each edge (black arrow): if the edge is oriented clockwise around a face - a triangle in this case - the purple vector in the middle of the edge points into the face, and otherwise it points out.  Note that this rule assumes an orientation on $M$.  Finally the vector field is extended into the interior of the face, resulting in at most an even singularity due to the Kasteleyn property, namely an odd number of clockwise pointing arrows around the face.}
\label{new_Kasteleyn}
\end{center}
\end{figure}

Another important fact is that both a (Kasteleyn orientation, dimer covering) pair and the corresponding spin structure naturally give rise to the same quadratic form on the $\Z_2$-valued homology $H_1(M,\Z_2)$ - see Theorem 4.1 of [\onlinecite{Kast}].  The utility of this result to us is that it implies that the Kasteleyn property holds not only on faces of $\Lambda$, but on certain other loops of $\Lambda$ as well.  Indeed, consider the difference $\Delta(D_1,D_2)$ between any two dimer coverings $D_1$ and $D_2$, i.e. the set of all dimers that are in only one of $D_1$ and $D_2$.  $\Delta(D_1,D_2)$ must consist of a disjoint union of loops, and Theorem 4.1 of [\onlinecite{Kast}] then guarantees that any such loop that bounds a disc - i.e. is trivial in $\Z_2$-valued homology of $M$ - must satisfy the Kasteleyn property; see figure \ref{new_central_plaquette_3}.  This property will be key in showing that the plaquette terms in the Hamiltonian constructed below conserve fermion parity.  Notice also that any such loop must have even length, so that it does not matter whether one is counting clockwise or counter-clockwise oriented edges.  Finally, we note that the Kasteleyn property will generally not hold on topologically non-trivial dimer difference loops.

Above we have described some very general properties of Kasteleyn orientations on arbitrary trivalent graphs and on surfaces with non-trivial topology.  It is worth emphasizing that if we are only interested in the ordinary planar hexagonal lattice, it is quite easy to pick an explicit Kasteleyn orientation on the associated graph $\Lambda$ which preserves much of the lattice symmetry, as illustrated in figure \ref{new_full_lattice}.  Indeed, for just the planar hexagonal lattice, the orientation where all the edges point from the `A' sublattice to the `B' sublattice in the standard bipartite decomposition is Kasteleyn: it has exactly 3 clockwise edges around each hexagonal plaquette.  Now, if we take this orientation for the type I edges in the graph $\Lambda$, and orient all of the type II edges in a clockwise direction around their respective triangles, then the result is a Kasteleyn orientation for $\Lambda$.  This Kasteleyn orientation preserves all of the translational symmetry and some of the rotational symmetry of $\Lambda$.

\section{Hamiltonian} \label{sec4}

The Hamiltonian of our model will be a sum of two terms
\begin{align}
H=H_{\text{fermion}} + H_{\text{fluct}},
\end{align}
The first term, $H_{\text{fermion}}$ will be defined to pick out a unique fermionic state for any configuration of spins.  Specifically, it picks out the fermionic state for which $i \gamma_v \gamma_w=1$ for every dimer $[vw]$ in the dimer covering associated to the spin configuration (see section \ref{sec2} above for the definition of the dimer covering associated with a spin configuration), where the edge $\la v w \ra$ is oriented from $v$ to $w$.  Formally, let
\begin{align}
D_{v w} = \frac{1-\tau^z_f \tau^z_{f'}}{2}
\end{align}
be the operator which detects a domain wall on edge $\la v w \ra$.  Here $v$ and $w$ are assumed to be nearest neighbors, and $f$ and $f'$ are the two faces which share the edge $\la v w \ra$.  Then set

\begin{figure}[htbp]
\begin{center}
\includegraphics[width=0.36\textwidth]{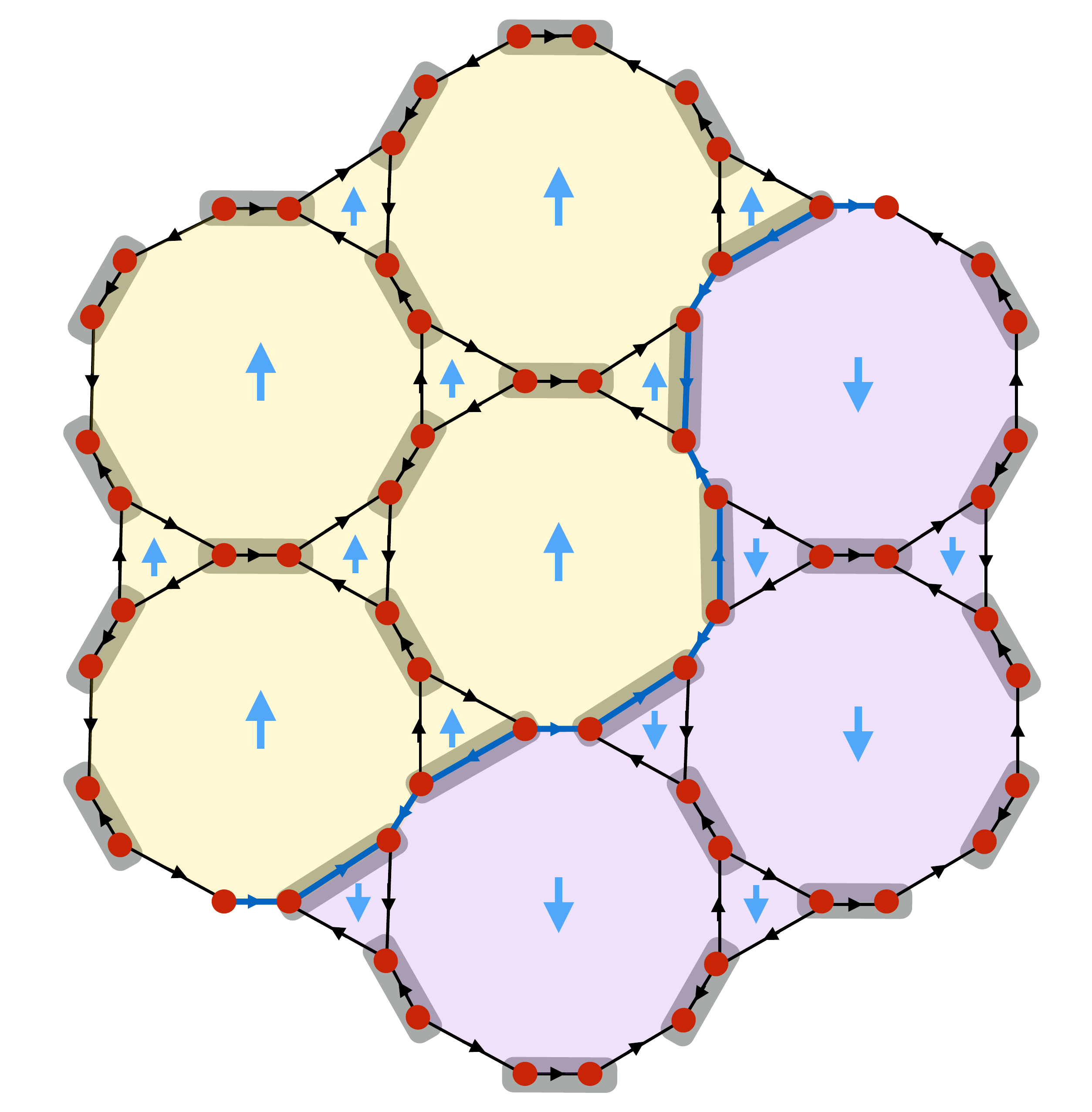}
\caption{Dimer covering of $\Lambda$ associated to a particular configuration of spins.  Away from the domain walls, dimers form across the type I edges, i.e. ones that connect different triangles, whereas along domain walls the dimers form on intra-triangular type II edges.  Note that the end of a domain wall carries an unpaired Majorana mode, illustrated here as an unpaired red dot.}
\label{new_domain_wall}
\end{center}
\end{figure}

\begin{figure*}[htbp]
\begin{center}
\includegraphics[width=0.8\textwidth]{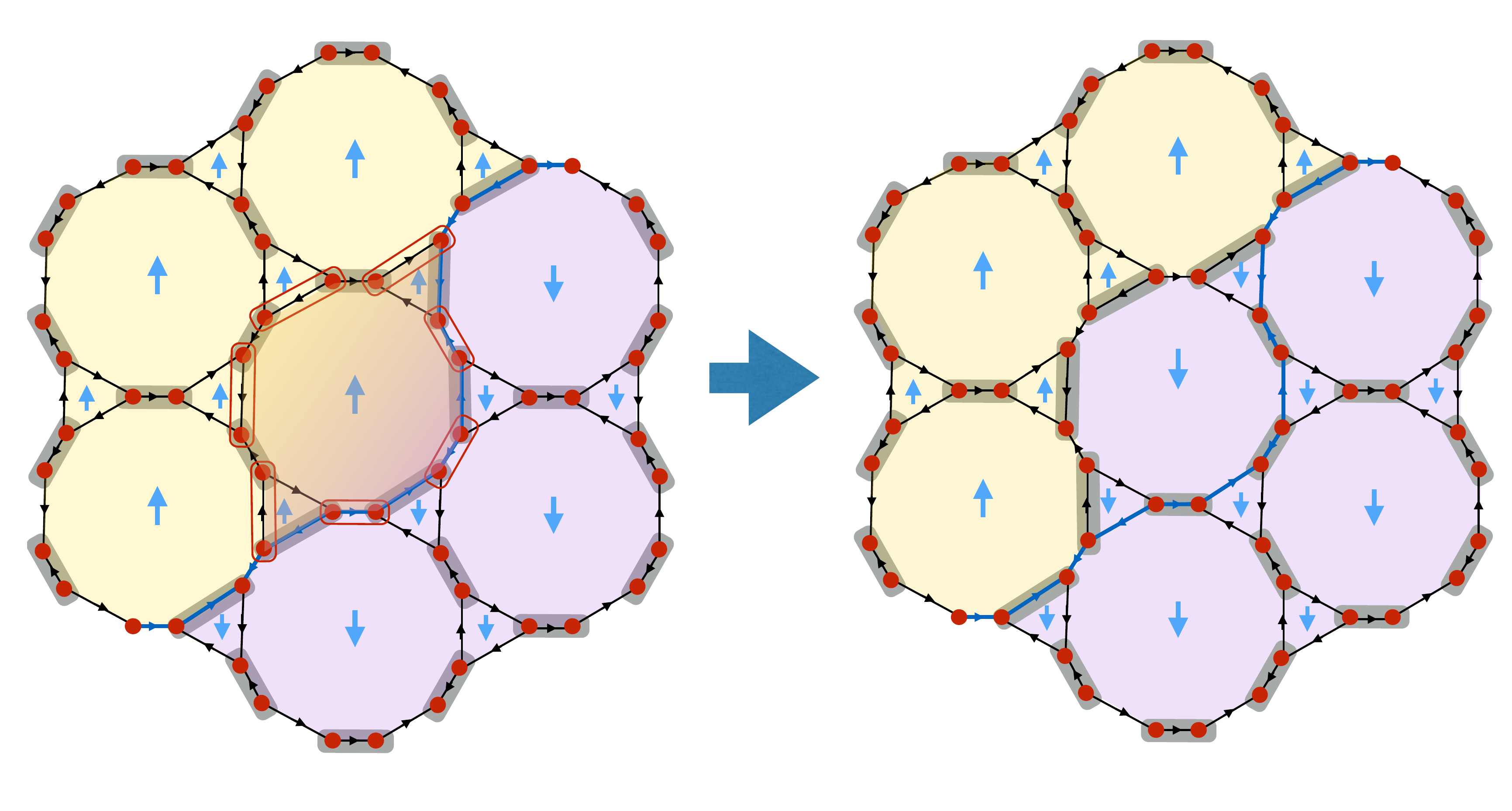}
\caption{Action of the `plaquette' term $\tau_p^x X_p$, defined in eq. \ref{Hfluct}, with $p$ being the central plaquette here.  This term acts on the fermions in such a way as to change the associated dimer coverings as indicated.  The difference between the two dimer coverings forms a short loop encircling $p$ and some of the its neighboring triangles, and $X_p$ essentially just projects on states with well defined fermion parity on each of the new dimers on this loop (rectangles outlined in red).  This parity is set by the Kasteleyn orientation.  As discussed in the text, the Kasteleyn property holds along this difference loop, which ensures that the fermionic states associated to the old and new dimer coverings have the same overall fermion parity.}
\label{new_central_plaquette_3}
\end{center}
\end{figure*}

\begin{align}
H_{\text{fermion}} &= -\sum_{\substack{\la \overrightarrow{vw} \ra \\ t(v)=t(w)}} i D_{vw} \gamma_v \gamma_w \\
  &- \sum_{\substack{\la \overrightarrow{vw}\ra\\ t(v) \neq t(w)}} i(1-D_{vw})\gamma_v \gamma_w
\end{align}
Here $\la \overrightarrow{vw}\ra$ means that the edge $\la v w \ra$ is oriented from $v$ to $w$.  The geometrical interpretation of this term is that it binds Majorana chains to domain walls, as illustrated in figure \ref{new_domain_wall}.  A key fact is that due to the Kasteleyn property, the fermionic states associated to any two global domain wall configurations have the same fermion parity.  Below we will prove this by showing that the individual `plaquette' terms which one applies to turn one such configuration into the other are all nonzero fermion parity even operators.  These `plaquette' terms are part of $H_{\text{fluct}}$, which we now define:

\begin{align} \label{Hfluct}
H_{\text{fluct}} = \sum_p \tau_p^x X_p
\end{align}
where $X_p$ rearranges the fermion configuration so that the Majorana chains follow the domain wall configurations as $\tau_p^x$ is applied.  Specifically,

\begin{align}\label{eq:decomp}
X_p =& \sum_{\substack{\{d_{vw}=0,1\} \\ \la v w \ra \in \partial p, t(v)\neq t(w)}} X_p^{\{d_{vw}\}} \Pi_p^{\{d_{vw}\}} P_p^{\{d_{vw}\}}.
\end{align}
Here the sum is over all $2^6=64$ possible domain wall configurations for the $6$ type I edges (i.e. those connecting different triangles) on the boundary $\partial p$ of the plaquette $p$.  Note that any one of these $64$ domain wall configurations determines also the domain wall configuration on all of the triangles bordering $p$ (see figure \ref{new_domain_wall_config}).  The operators $P_p^{\{d_{vw}\}}$ and $\Pi_p^{\{d_{vw}\}}$ are projectors: $P_p^{\{d_{vw}\}}$ projects onto spin states that have domain walls precisely where $d_{vw}=1$ (i.e. states with $D_{vw}=d_{vw}$), and $\Pi_p^{\{d_{vw}\}}$ projects onto states in the fermionic Hilbert space that conform to those domain walls:

\begin{figure}[htbp]
\begin{center}
\includegraphics[width=0.4\textwidth]{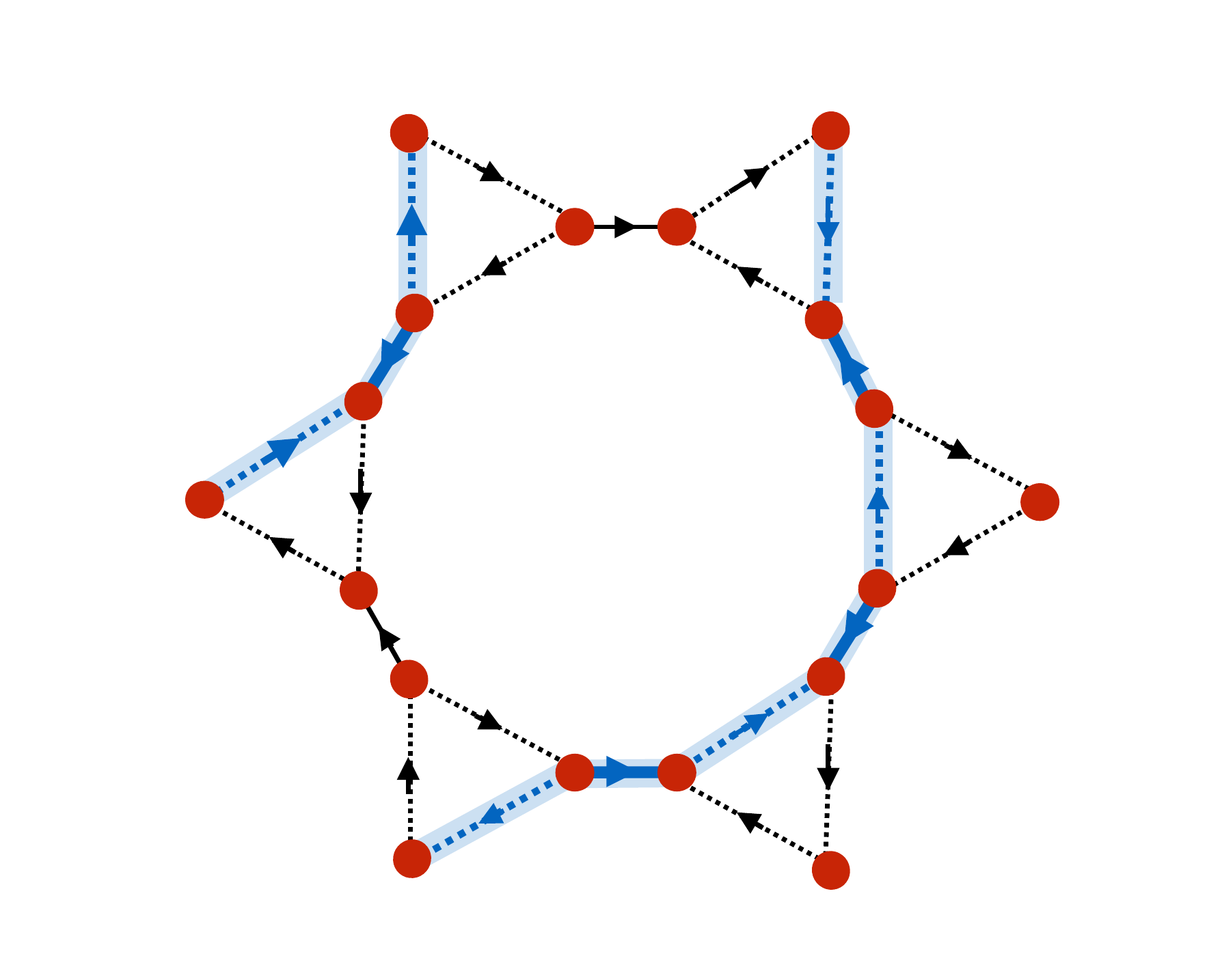}
\caption{Domain walls in the neighbourhood of a plaquette $p$.  Solid lines indicate type I edges, dotted lines type II edges, and blue denotes domain walls.  There are $6$ type I edges and thus $2^6=64$ possibilities $\{d_{vw} =0,1\}$ for domain wall configurations on such edges.  Using the `majority rule' extension of spin configurations to triangles, each such possibility determines uniquely the domain wall configuration on all of the adjoining triangles as well.  This information in turn determines uniquely the difference loop between new and old dimer configurations, on which the projectors in $X^{\{d_{vw}\}}$ act.}\label{new_domain_wall_config}
\end{center}
\end{figure}

\begin{align}
P_p^{\{d_{vw}\}} = \prod_{\substack{\la v w \ra \in \partial p \\ t(v)\neq t(w)}}\left( \frac{1+(-1)^{d_{vw}-D_{vw}}}{2} \right)
\end{align}

\begin{align}
\Pi_p^{\{d_{vw}\}} =& \prod_{ \substack{\la \overrightarrow{vw} \ra \in \partial'p \\ t(v)=t(w), d_{vw}=1}} \left(\frac{1-i\gamma_v \gamma_w}{2}\right) \cdot \\
& \prod_{ \substack{\la \overrightarrow{vw} \ra \in \partial'p \\ t(v)\neq(w), d_{vw}=0}} \left(\frac{1-i \gamma_v \gamma_w}{2}\right)
\end{align}
Here $\partial'p$ is the set of $18$ vertices in the triangles surrounding $p$.  

\subsection{Definition of plaquette term $X_p^{\{d_{vw}\}}$}\label{ssec:Xpdef}

We now define the action of $X_p^{\{d_{vw}\}}$ on the fermionic degrees of freedom for a specific domain wall configuration $\{d_{vw}=0,1\}$ in the vicinity of plaquette $p$, completing the definition of the Hamiltonian.  It is best to think in terms of dimer coverings.  The fermionic state being acted upon corresponds to some initial dimer covering $D_i$, and acting with $X_p^{\{d_{vw}\}}$ results in a state corresponding to a final dimer covering $D_f$.  The difference $\Delta(D_i,D_f)$ - that is, the set of edges which form dimers in precisely one of $D_i$ and $D_f$ - forms a loop around the face $p$ together with some of the adjoining triangles - see figure \ref{new_central_plaquette_3}.  Let us denote the vertices along this loop by $v_1, \ldots, v_{2n}$, and for simplicity let us change the notation from $\gamma_{v_j}$ to $\gamma_j$.  Suppose $[12],[34],\ldots,[2n-1,2n]$ form dimers in $D_i$.  Then $X_p^{\{d_{vw}\}}$ must change $[12],[34],\ldots,[2n-1,2n]$ to $[23],[45],\ldots,[2n,1]$.  With this in mind, define

\begin{align} \label{Xpdef}
X_p^{\{d_{vw}\}} =& 2^{-\frac{n+1}{2}} (1+is_{2,3} \gamma_{2}\gamma_{3})\ldots (1+is_{2n,1} \gamma_{2n} \gamma_{1})
\end{align}
Here we define $s_{i,j}=1$ if the edge $\la v_i v_j \ra$ is oriented from $v_i$ to $v_j$, and $s_{i,j}=-1$ otherwise.  Since the right side of eq. \ref{Xpdef} is a product of the appropriate projectors, it is clear that acting with $X_p^{\{d_{vw}\}}$ results in a fermionic state corresponding to $D_f$, but we have to check that this state is in fact nonzero, for otherwise such fluctuations would not occur.  We will in fact see that this state has norm $1$.

To show this, we can work in the reduced Fock space of the Majoranas $\gamma_1, \ldots, \gamma_{2n}$.  Let $|\psi \ra$ be the state in this reduced Fock space defined uniquely up to phase by the condition $is_{2j-1,2j} \gamma_{2j-1} \gamma_{2j}=-1$ for $j = 1,\ldots, n$.  Then we just have to demonstrate that 

\begin{align} \label{chipsi}
|\chi\ra\equiv2^{-\frac{n+1}{2}} (1+is_{2,3} \gamma_2 \gamma_3) \ldots (1+is_{2n,1} \gamma_{2n} \gamma_1) |\psi\ra
\end{align}
has norm $1$.  We now make use of the Kasteleyn property, which implies that the number of clockwise arrows along our loop is odd, as discussed in the previous subsection (since our loop is a topologically trivial difference loop between two dimer configurations).  The Kasteleyn property is key: had it been violated, we would have had $|\chi\ra=0$, and the states corresponding to the two dimer coverings would have differed in fermion parity.

To prove that $|\chi\ra$ defined in eq. \ref{chipsi} has norm $1$, first expand the product in equation \ref{chipsi}.  This results in a sum of $2^n$ terms $|\psi_l\ra$, $l=1,\ldots, 2^n$.  They can be paired up into $2^{n-1}$ pairs as follows: $|\psi_l\ra$ is paired with $|\psi_{l'}\ra$ if they differ in the choice made in each of the $n$ factors being expanded.  In this case, using the fact that $i\gamma \gamma'$ squares to $1$ for $\gamma \neq \gamma'$, we have:

\begin{align}
|\psi_{l'}\ra=(is_{2,3} \gamma_2 \gamma_3)\ldots (is_{2n,1} \gamma_{2n} \gamma_1) |\psi_l\ra
\end{align}
Now, the Kasteleyn condition reads $s_{1,2} s_{2,3} \ldots s_{2n-1,2n} s_{2n,1}=-1$ (since the loop has even length, this is independent of whether we go clockwise or counter-clockwise around the loop).  Also,

\begin{align}
\gamma_2 \gamma_3 \ldots \gamma_{2n} \gamma_1 = -\gamma_1 \ldots \gamma_{2n}
\end{align}
Using these two facts, we see that
\begin{align} \label{middleEq}
|\psi_{l'}\ra&=(is_{1,2} \gamma_1 \gamma_2) \ldots (is_{2n-1,2n} \gamma_{2n-1} \gamma_{2n}) |\psi_l\ra \\
&=|\psi_{l}\ra
\end{align}
with the last equality following from the fact that $(is_{2j-1,2j} \gamma_{2j-1} \gamma_{2j})|\psi\ra=|\psi\ra$, and the fact that all of the $(is_{2j-1,2j} \gamma_{2j-1} \gamma_{2j})$ terms in equation \ref{middleEq} can be commuted past the various $\gamma$ bilinears that appear in the definition of $|\psi_l\ra$ in terms of $|\psi\ra$.  Thus $|\psi_l\ra = |\psi_{l'}\ra$ for each pair $|\psi_l\ra, |\psi_{l'}\ra$.  Notice that if the Kasteleyn condition had been violated, these two would be negatives of each other and the resulting state $|\chi\ra$ would have been $0$.

Furthermore, it is easy to see, for example by examining the fermion occupation numbers in the basis corresponding to pairing up $\gamma_{2j-1}$ with $\gamma_{2j}$ for $j=1,\ldots, n$, that if $l$ and $k$ are not in the same pair then $|\psi_l\ra$ and $|\psi_k\ra$ are orthogonal.  Therefore

\begin{align}
||\chi\ra|^2 = 2^{n-1} \cdot \left( 2\cdot 2^{-\frac{n+1}{2}} \right)^2 = 1
\end{align}
as desired.  Note that in particular, we have shown that the fermionic states associated to any two domain configurations must have the same fermion parity, since they can be connected to each other by a sequence of such plaquette moves.

\section{Commutation relations of $\tau^x_p X_p$} \label{sec_comm}

The plaquette operators ($\tau^x_p X_p$) defined above clearly commute up to a domain wall dependent phase factor.  In this somewhat technical section, we show that in fact, these operators commute exactly, as required in a commuting projector model.  Readers who can take this on faith can skip this section.

First, let us examine how `local' a single plaquete operator is, i.e. which nearby spins and fermionic degrees of freedom it acts on.  These will form a Hilbert space $\mathcal{H}^{\text{local}}$, while the complementary degrees of freedom form $\mathcal{H}^{\text{other}}$, with the tensor product of these two being the whole Hilbert space $\mathcal{H}$.  Consulting the definition of the plaquette operator (equation \ref{eq:decomp}), we see immediately that it acts on the Ising spin as well as the Fock space of the 18 Majorana operators surrounding the plaquette.  We will refer to this as the \emph{neighbourhood} of $p$ (figure \ref{new_domain_wall_config}).

It is clear that any two non-adjacent plaquette operators commute, simply because their neighborhoods do not overlap and the plaquette operators are fermion parity even.  The non-trivial case thus consists of adjacent plaquettes.  In this case, we can still factor their action on the Ising spins, but not on the fermionic Fock space. This is because the neighbourhoods of the plaquettes share Majoranas, so we cannot disentangle the action into pieces that act exclusively in one of the two regions.  However, we can still ignore the region away from the union of the neighbourhoods, since both plaquettes act as the identity there, and just focus on the neighbourhood of the two adjacent plaquettes.  As illustrated in figure \ref{Plaquette_support}, the appropriate local Fock space is formed by the 30 Majoranas in this neighbourhoods of $p_1$ and $p_2$.  From this point forward, we work exclusively with this local Fock space.

\begin{figure}[htbp]
\begin{center}
\includegraphics[width=0.4\textwidth]{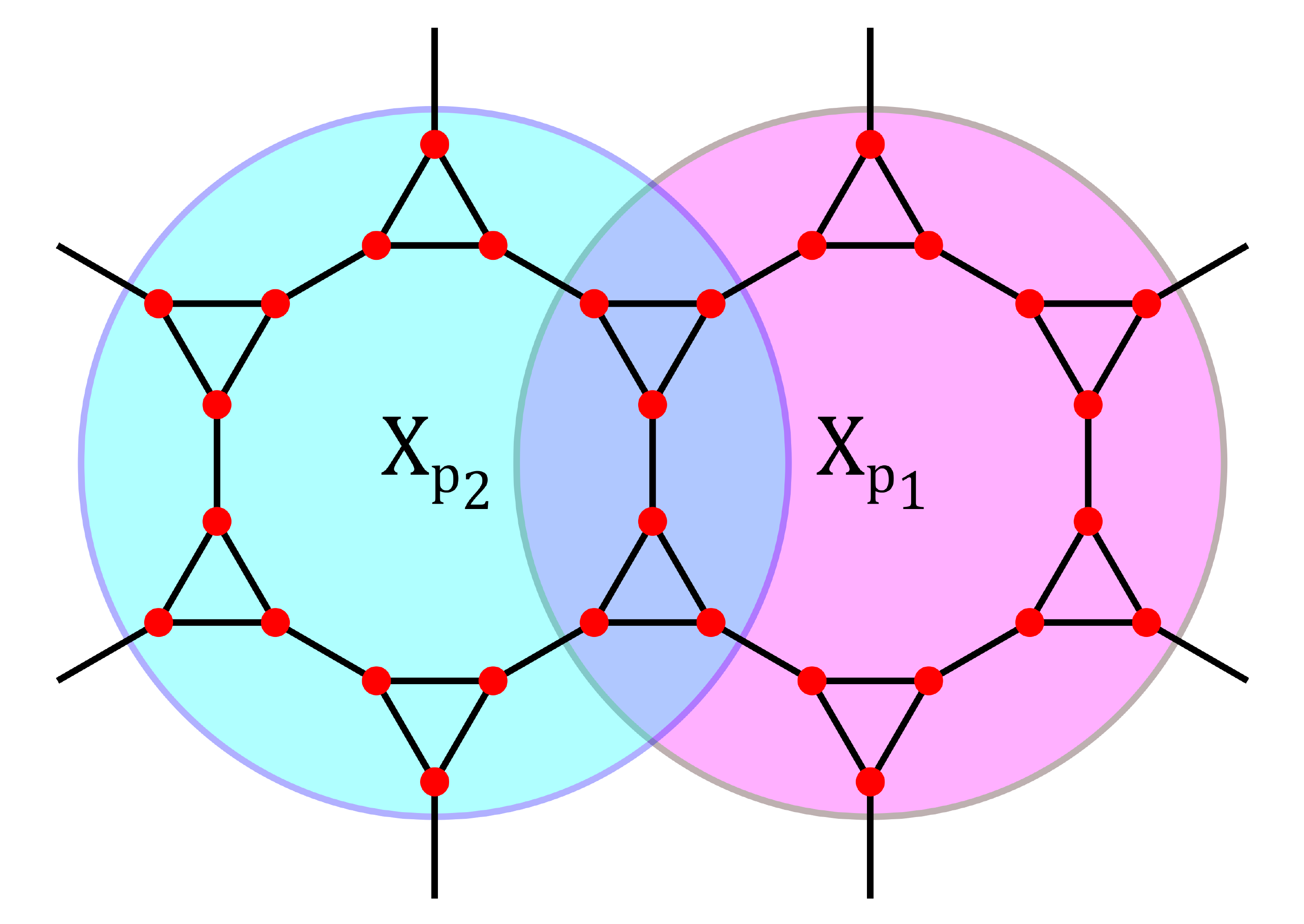}
\caption{Majoranas in the neighbourhood of $p_1$ (magenta region) and $p_2$ (cyan region). The plaquette operators modify dimers within either neighbourhood, but leave dimers outside untouched. Since the two regions share Majoranas (purple intersection), we cannot factor the Fock space further.}
\label{Plaquette_support}
\end{center}
\end{figure}

We can reduce the size of the space we need to consider further. Looking back to the definition of $X_p$ in equation \ref{eq:decomp}, we note that it is decomposed into 3 pieces: $X^{\{d_{vw}\}}_{p}$, $\Pi^{\{d_{vw}\}}_{p}$ and $P^{\{d_{vw}\}}_{p}$. The latter two of these are explicitly projectors. Specifically, $\Pi^{\{d_{vw}\}}_{p}$ projects onto the fermionic state which conforms to the domain walls fixed by $P^{\{d_{vw}\}}_{p}$. This means that we need only consider the action of $X_{p_1}$ and $X_{p_2}$ on states where the fermionic and domain wall data match, which we write as $\ket{\Psi(\{d_{vw}\})}$. Any other state will be annihilated by these projectors.

In summary, to show that adjacent plaquette terms commute, we need only show that they do so on states of the form $\ket{\Psi(\{d_{vw}\})}\otimes\ket{\tau_1,\tau_2}$, where $\tau_1$ and $\tau_2$ are the 2 Ising spins being acted on, and $\ket{\Psi(\{d_{vw}\})}$ is the fermion many-body state conforming to the domain wall configuration $\{d_{vw}\}$ on the restricted, 30 Majorana lattice. On such a state, the plaquette term action simplifies to
\begin{align}
&\tau^x_{p_1}X_{p_1}\ket{\Psi(\{d_{vw}\})}\otimes\ket{\tau_1,\tau_2}\nonumber\\
&= X^{\{d_{vw}\}}_{p_1}\ket{\Psi(\{d_{vw}\})}\otimes\ket{\tau'_1,\tau_2}
\label{plaq12}
\end{align}
as $\ket{\Psi(\{d_{vw}\})}$ is left unchanged by $\Pi_{p_1}^{\{d_{vw}\}}$.

\begin{figure}[htbp]
\begin{center}
\includegraphics[width=0.4\textwidth]{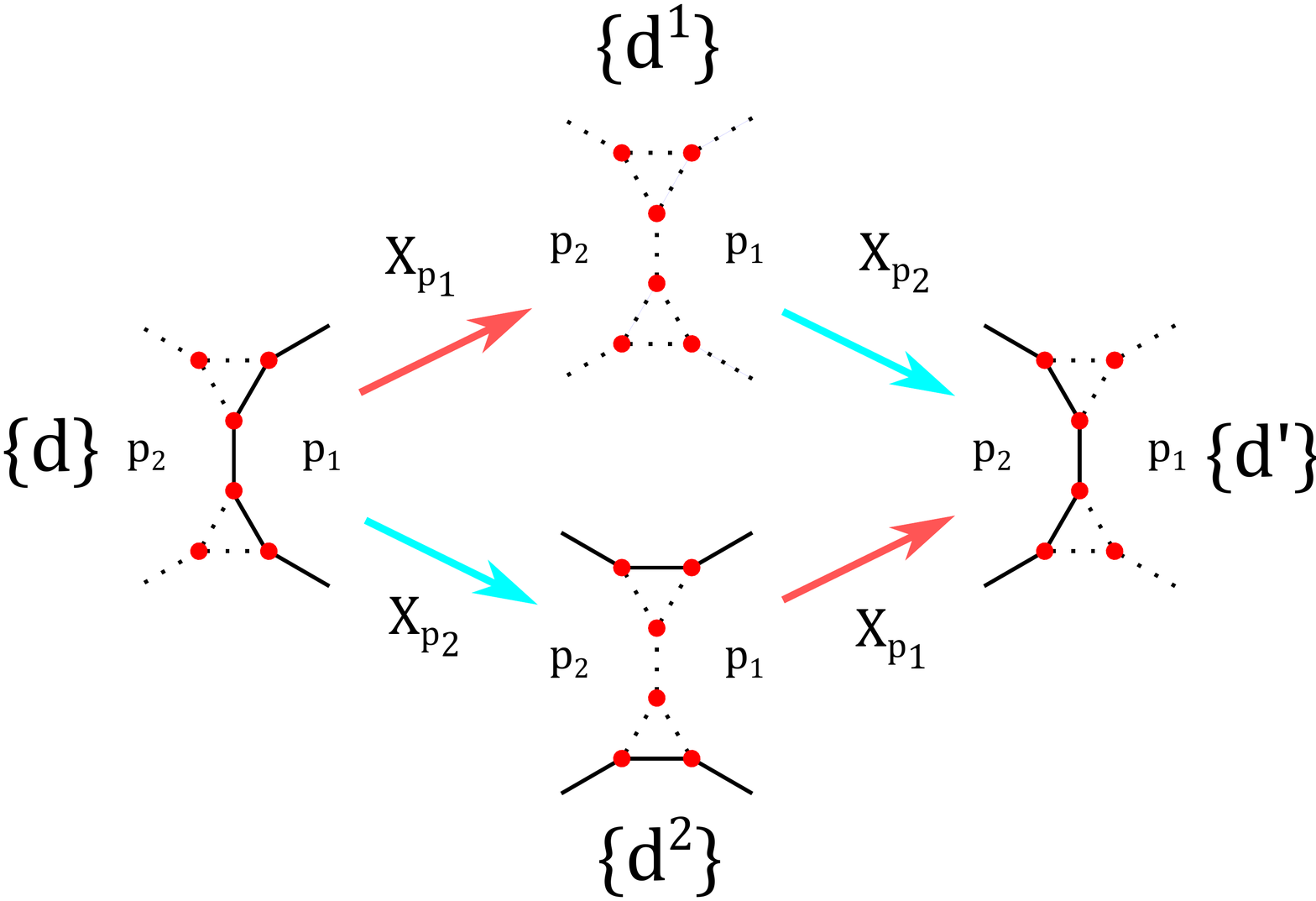}
\caption{Illustration of the 2 possible domain wall fluctuation paths, viewed on the shared neighbourhood of $p_1$ and $p_2$. Solid lines mark domain walls, while dashed lines are present for clarity only. Starting from a domain wall configuration $\{d_{vw}\}$, we move to either $\{d^1_{vw}\}$ or $\{d^2_{vw}\}$, depending on which plaquette operator is used first. The final domain wall configuration, $\{d'_{vw}\}$, does not depend on the order of operation. }
\label{Plaquette_comdiag}
\end{center}
\end{figure}

Acting with a plaquette term modifies an Ising spin, which in turn changes the domain wall configuration. Starting from an initial configuration $\{d_{vw}\}$, acting $\tau^x_{p_1}X_{p_1}$ and $\tau^x_{p_2}X_{p_2}$ will take us to some final configuration $\{d'_{vw}\}$, which does not depend on the order of operation. However, the intermediate arrangement must obviously depend on which plaquette operator acts first (figure \ref{Plaquette_comdiag}). We need to keep track of this, as the behaviour of $X_p$ depends on the domain wall configuration it acts upon.

\begin{figure}[htbp]
\begin{center}
\includegraphics[width=0.45\textwidth]{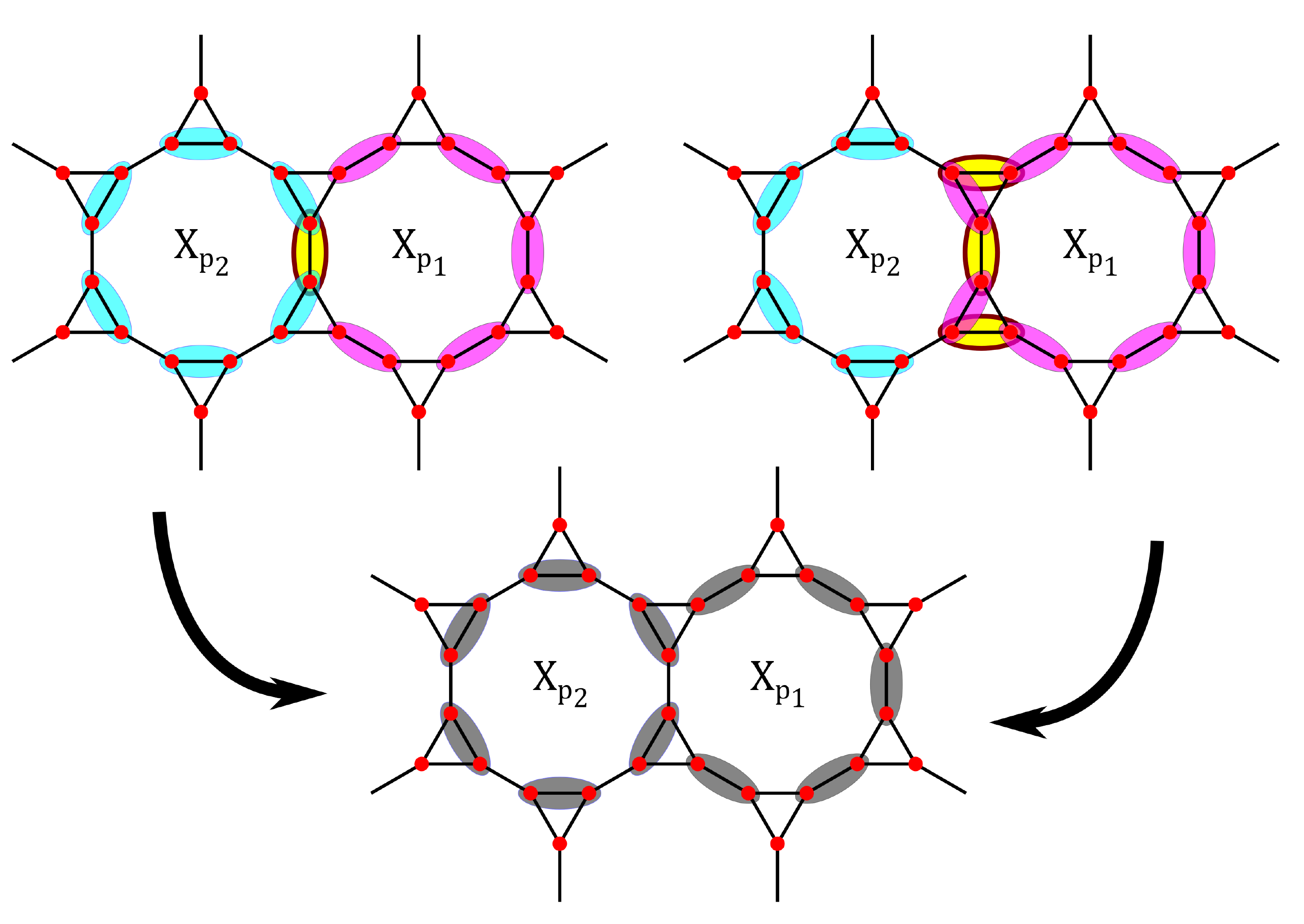}
\caption{Digramatic representation of the projectors (magenta, cyan and yellow ovals) involved in equations \ref{plaq12} (top left) and \ref{plaq21} (top right). The projectors fail to commute are highlighted in yellow. The remaining projectors (gray ovals, bottom) form a commuting set of operators. }
\label{plaq_equiv}
\end{center}
\end{figure}

Acting $\tau^x_{p_1}X_{p_1}$ first, we will pass via a domain configuration $\{d^1_{vw}\}$ before reaching the final state, and so the action of both plaquette operators on the state is
\begin{align}
&\tau^x_{p_2}X_{p_2}\tau^x_{p_1}X_{p_1}\ket{\Psi(\{d_{vw}\})}\otimes\ket{\tau_1,\tau_2}\nonumber\\
&= X^{\{d^1_{vw}\}}_{p_2}X^{\{d_{vw}\}}_{p_1}\ket{\Psi(\{d_{vw}\})}\otimes\ket{\tau'_1,\tau'_2}
\label{plaq12}
\end{align}
while acting $\tau^x_{p_2}X_{p_2}$ takes us through a configuration $\{d^2_{vw}\}$, producing an a priori different result.
\begin{align}
&\tau^x_{p_1}X_{p_1}\tau^x_{p_2}X_{p_2}\ket{\Psi(\{d_{vw}\})}\otimes\ket{\tau_1,\tau_2}\nonumber\\
&= X^{\{d^2_{vw}\}}_{p_1}X^{\{d_{vw}\}}_{p_2}\ket{\Psi(\{d_{vw}\})}\otimes\ket{\tau'_1,\tau'_2}
\label{plaq21}
\end{align}
We have now reached a purely algebraic question: Are the right-hand sides of equations \ref{plaq12} and \ref{plaq21} equal? Recalling the definition of $X^{\{d_{vw}\}}_{p}$ in equation \ref{Xpdef}, we see that they are products of projectors, ie terms of the form $(1+i \gamma_v\gamma_w)$. This fixes the parity of $i \gamma_v\gamma_w$ to be 1. These projectors occasionally fail to commute. Since they are built from Majorana bilinears, we should check to see the conditions under which bilinears fail to commute. Two such operators, $i \gamma_v\gamma_w$ and $i \gamma_x\gamma_y$, \emph{anticommute} if they share exactly one Majorana. The corresponding projectors fail to commute in that case.

If we attempt to reorder the projectors in $X^{\{d^1_{vw}\}}_{p_2}X^{\{d_{vw}\}}_{p_1}$ and $X^{\{d^2_{vw}\}}_{p_1}X^{\{d_{vw}\}}_{p_2}$ to make them identical, we can almost succeed. Our process is obstructed by a number of non-commuting projectors, all of which are built with Majoranas from the intersection between plaquette neighbourhoods (yellow in figure \ref{plaq_equiv}). This makes intuitive sense, since the action of each plaquette term is localized to its neighbourhood, so only the intersection ``sees" the order of operation. If we could delete a few of these projectors from the product, we would be left with a commuting set of operators, represented in gray in figure \ref{plaq_equiv}. 

It is useful to think of each projector as a constraint on our Fock space, since each enforces that our state live in some subspace. If two projectors commute, the corresponding operator constraints can both be satisfied simultaneously. Our commuting set of projectors can be thought of then as constraints which, when satisfied, completely fix the state. The projectors which fail to commute are the ones which risk making the state over-constrained. Thankfully, these projectors give constraints on the state which are already satisfied, provided that the dimer orientations satisfy the Kasteleyn condition. Thus, the product of these non-commuting projectors is guaranteed to act as the identity on our state, and so they can be ``absorbed" (treated like the identity).

Proving the aforementioned statement involves a lengthy sequence of projector identities, and we do this in appendix \ref{append}.

\section{Analysis of SPT order and generalization to arbitrary fermionic SPTs} \label{sec_analysis}

\subsection{Analysis of SPT order}
We claim that the fermionic Hamiltonian constructed above has the following properties: 

{\bf (1)} Upon breaking the $\Z_2$ symmetry, it can be continuously connected to a trivial fermionic insulator without closing the gap.

{\bf (2)} A $\Z_2$ symmetry flux traps a Majorana zero mode.

{\bf (3)} The local $\Z_2$ symmetry action changes the fermion parity bound to a $\pi$ flux.

Property (1) establishes that we have a 2+1d fermionic SPT of an onsite unitary $\Z_2$ symmetry.  It is known that the integer free fermion classification of SPTs in this symmetry class is broken down to $\Z_8$ by interactions\cite{GuLevin, Qi_Z8, Ryu_Z8, Yao_Z8}, and furthermore it has been conjectured\cite{Kapustin2015} that there are no other interacting fermionic SPTs in this symmetry class, i.e. the full interacting classification is $\Z_8$.  Under this assumption, either of properties (2) or (3) above establishes that our Hamiltonian represents one of the odd $\nu \in \Z_8$ phases.  Indeed, these properties are easy to see in the free fermion model of an SPT with index $\nu$, namely a $p+ip$ superconductor with Chern index $\nu$ stacked with a $p-ip$ superconductor with Chern index $-\nu$.  In this model the $\Z_2$ symmetry just measures the fermion parity of the $p+ip$ layer, and a symmetry flux is simply a $\pi$ flux in only this $p+ip$ layer.  This symmetry flux thus binds a Majorana zero mode precisely when $\nu$ is odd.  A similar argument shows that acting with the $\Z_2$ symmetry in the vicinity of a $\pi$ flux that penetrates both layers must change the fermion parity at such a $\pi$ flux.

Let us now demonstrate properties (1)-(3).  The key to seeing (1) is that making the terms in $H_{\text{fermion}}$ large, i.e. giving a large energetic penalty to fermionic configurations that do not conform to a given domain wall configuration, yields a low energy Hilbert space that can be mapped to the Ising model.  Indeed, the $\tau$ spin configuration uniquely determines, up to phase, the state in this low energy Hilbert space.  The relative phase between any two such states can be fixed by demanding that one be sent to the other by a sequence of plaquette operators $X_p$; the fact that the plaquette operators commute and all square to $1$ means that this definition is independent of the sequence chosen.  In such a basis for the low energy Hilbert space the effective Hamiltonian is just that of the trivial Ising paramagnet, with a transverse field but with zero Ising coupling.  In such an Ising model of decoupled spins, breaking the spin flip symmetry and turning on a field in the $z$ direction, while simultaneously turning off the transverse field, continuously connects the paramagnet to a fully polarized ferromagnet, without opening a gap.  Similarly, in our model we can turn on the operator $\tau_p^z$, dressed by a projector onto the low energy Hilbert space, and continuously deform to a model whose ground state has no domain walls, and all link fermions empty, i.e. a trivial fermionic insulator.

Property (2) follows from the way the fermions are bound to domain walls.  Indeed, from figure \ref{new_domain_wall} it is clear that a domain wall endpoint binds an unpaired Majorana mode.  To see property (3), note that a pair of $\pi$ fluxes at plaquettes $p_1$ and $p_2$ can be inserted in our model by reversing the orientation of a sequence of edges crossed by a path from $p_1$ to $p_2$.  As a result, the Kasteleyn condition is violated at $p_1$ and $p_2$, so the corresponding plaquette terms $X_{p_1}$ and $X_{p_2}$ act as $0$.  By multiplying $X_{p_1}$ and $X_{p_2}$ by appropriate Majorana operators, it is possible to construct fermion parity odd operators that act locally with the $\Z_2$ symmetry near $p_1$ and $p_2$ respectively.

\subsection{Generalization to arbitrary fermionic SPTs}

Consider a general symmetry group $G_f = G \times \Z_2^f$, with $G$ finite.  According to [\onlinecite{M_Cheng}], fermionic SPTs are classified by $3$ pieces of data.  The first is a group map $\sigma: G \rightarrow \Z_2$, which tells us whether a the symmetry flux of a particular $g\in G$ traps a Majorana zero mode.  Any two fermionic SPTs with the same $\sigma$ must differ by stacking a group-supercohomology model according to [\onlinecite{M_Cheng}], and such group-supercohomology models are known to have commuting projector representations.  Thus, in order to show that all 2+1d fermionic SPTs have commuting projector realizations, we only need to construct one such model for each choice of $\sigma$.  

But such a construction is a trivial generalization of our $\Z_2$ model: we simply construct a model with a $|G|$ dimensional spin on each plaquette, and define a $\Z_2$ domain wall to exist between $f$ and $g$ if $\sigma(f^{-1} g) \neq 0$.  The Hamiltonian then binds fermionic configurations to these $\Z_2$ domain walls as before.  The plaquette terms allow fluctuations from any $g$ to any other $g'$ on a given plaquette, with the stipulation that if $\sigma(g^{-1} g') \neq 0$, then the plaquette term rearranges the fermionic configuration as discussed in the $\Z_2$ case above.

\section{Conclusions and future directions} \label{sec_future}

We have shown that all known fermionic SPTs in 2+1 dimensions have lattice Hamiltonian representations via commuting projectors, and furthermore can be put on 2d oriented surfaces $M$ of arbitrary topology.  We also showed that putting our models on such surfaces necessitates a choice of spin structure, manifesting in our construction as a choice of Kasteleyn orientation of an associated graph, whose vertices are a Majorana representation of the fermionic degrees of freedom.  There are several potential avenues for further investigation.

One is to relate our Kasteleyn version of a spin structure to that of reference [\onlinecite{G_K}].  We do not expect a direct connection, because the vertices in our graph represents Majorana zero modes, while the triangulations involved in reference [\onlinecite{G_K}] involve physical fermions.  Nevertheless, it would be good to put these two constructions on the same footing.  Another direction would be to generalize to anti-unitary symmetries like time reversal and continuous ones like $U(1)$, as well as symmetry groups which do not factor as a simple product of $\Z_2^f$ and a remaining piece.  We could then hope to access commuting projector Hamiltonians for more interesting 2+1d fermionic SPTs, such as the quantum spin Hall phase.  Yet another natural generalization would be to understand discrete versions of spin structures in 3+1 dimensions.

Besides SPTs, our work may have applications to commuting projector models of fermionic topological orders, which also require a spin structure.  It would be interesting to see whether the discrete versions of spin structures discussed here enter naturally into fermionic versions of string net models \cite{Gu_Wang_Wen}.  Finally, it would be good to relate this work to that of K. Walker \cite{Walker}, who gave a prescription for constructing fermionic Hamiltonians by starting with bosonic ones that contain an emergent fermion, and then `ungauging' fermion parity symmetry by condensing a bound state of the emergent fermion and an additional fundamental fermion degree of freedom.  The latter requires a choice of spin structure, which manifests itself in the phases of the various terms in the Hamiltonian that condense the bound state.  A particular instance of this construction appears to yield an exactly solved model in the same phase as the $\Z_2$ gauged version of an odd $\nu$ fermionic SPT, and it would be interesting to compare this construction with ours.

\section{Acknowledgements}

We thank John Morgan, Michael Levin, Chris Heinrich, Fiona Burnell, David Aasen, Ashvin Vishwanath, Andrew Potter, and Anton Kapustin for useful discussions.  Lukasz Fidkowski is supported by NSF grant DMR-1519579 and by the Alfred P. Sloan foundation.

\begin{appendix}
\section{Proof of the required projector identities} \label{append}

As noted in Sec. \ref{ssec:Xpdef}, $X^{\{d_{vw}\}}_{p}$ contains Majoranas that live on a loop on the graph $\Lambda$. It stands to reason then that 2 adjacent plaquette terms live on 2 loops which intersect along some edges. With this in mind, let us begin with 2 loops on our dimer graph, on which we have defined orientations $s_{vw}$. Each vertex comes with a Majorana, labeled $\gamma_i$ ($i\in [1\dots 2n]$) and $\gamma'_j$ ($j\in [1\dots 2m]$). On the Fock space of these Majoranas, we define operators
\begin{align}
\Gamma  &=2^{-\frac{n+1}{2}} \prod_{j=1}^{n}\left( 1+s_{2j,2j+1}i\gamma_{2j} \gamma_{2j+1}\right)\label{Gamma}\\
\Gamma' &= 2^{-\frac{m+1}{2}} \prod_{j=1}^{m}\left( 1+ s_{2j-1,2j}i\gamma'_{2j-1} \gamma'_{2j}\right)\label{Gamma'}\\
P &=  2^{-m} \prod_{j=1}^{m}\left( 1+ s_{2j,2j+1}i\gamma'_{2j} \gamma'_{2j+1}\right)\label{Proj}
\end{align}
Suppose now that these loop share some edges. This would correspond to identifying some Majoranas along the loop, so set $\gamma_i = \gamma'_i$ for $i \in  [1\dots 2k]$. $\Gamma$ and $\Gamma'$ are both operators with the same structure as $X^{\{d_{vw}\}}_{p}$. By appropriately choosing the size of the loop and location of the Majoranas, we can make $\Gamma$ and $\Gamma'$ into whatever $X^{\{d_{vw}\}}_{p}$ we choose. The operator $P$ is built so that it projects onto the subspace spanned by the fermion state $\ket{\Psi(\{d_{vw}\})}$. Writing this projector explicitly will allow us to exploit projector identities when simplifying later expressions. Taken together, the product $\Gamma\Gamma' P$ can, with some extra data, represent either equation \ref{plaq12} or \ref{plaq21}.

\begin{figure}[htbp]
\begin{center}
\includegraphics[width=0.4\textwidth]{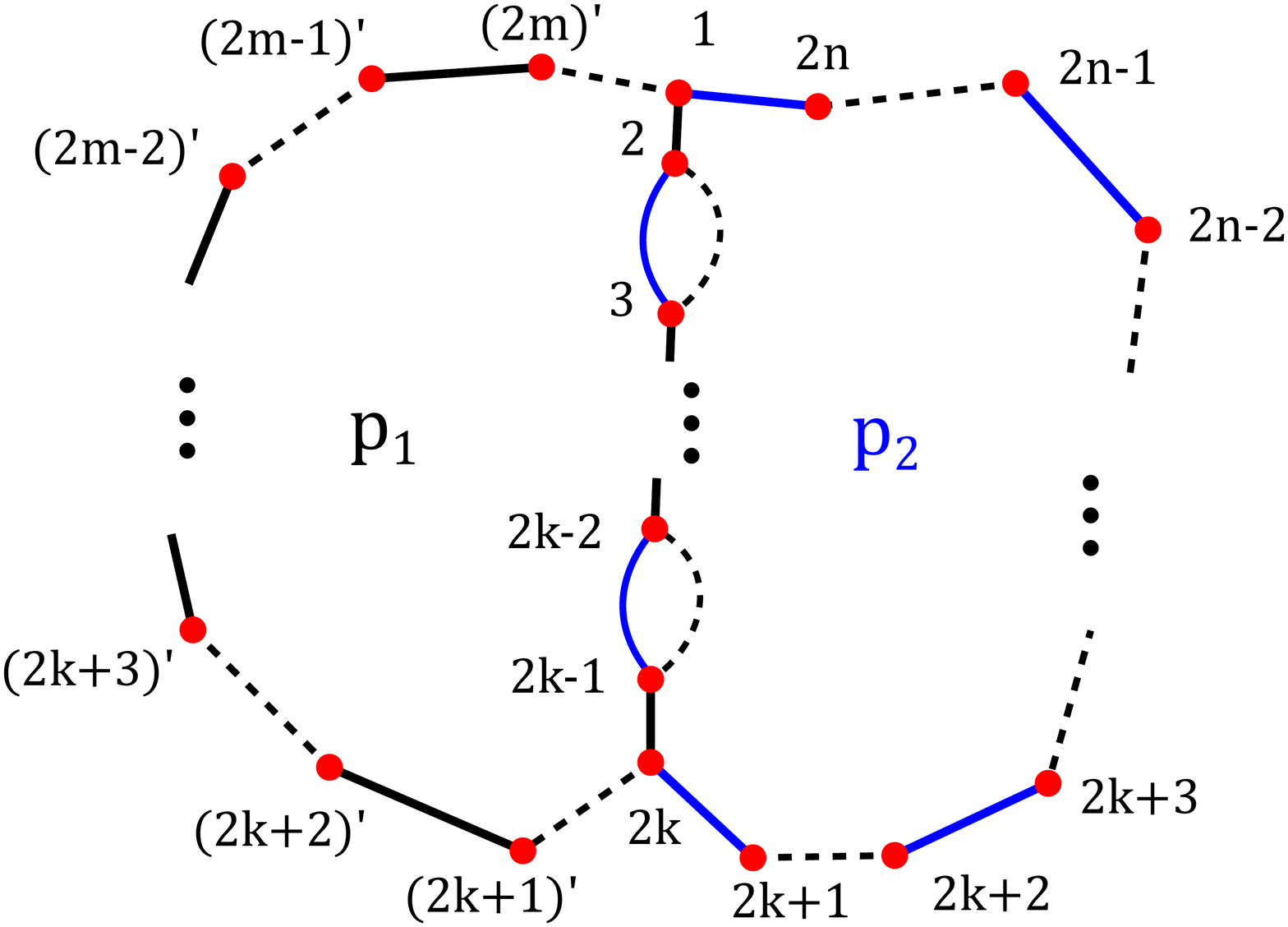}
\caption{Representation of all the Majoranas contained in $\Gamma$, $\Gamma'$ and $P$. Individual projectors are represented by links, with blue being those in $\Gamma$ , solid black in $\Gamma'$  and dashed black being those in $P$.}
\label{2plaquettes}
\end{center}
\end{figure}

There are many projectors to keep track of, but each of them corresponds to a dimer, and so we represent them pictorially in figure \ref{2plaquettes}. We see immediately that we can split our vertices into 2 classes: Vertices where $\Gamma$, $\Gamma'$ and $P$ act (the shared edge of the loops), and vertices where either $\Gamma$ or $\Gamma'$ act alongside $P$ (the outer edge of both loops). It is reasonable then to split our operators into two parts, which we call interior (on the shared edge) and exterior (on the outer edges).
\begin{align}
\Gamma_{int} &=2^{-\frac{k-1}{2}} \prod_{j=1}^{k-1}\left( 1+i s_{2j,2j+1}\gamma_{2j} \gamma_{2j+1}\right)\\
\Gamma_{ext} &=2^{-\frac{n-k+2}{2}} \prod_{j=k}^{n}\left( 1+i s_{2j,2j+1}\gamma_{2j} \gamma_{2j+1}\right)\\
\Gamma'_{int} &= 2^{-\frac{k}{2}} \prod_{j=1}^{k}\left( 1+i s_{2j-1,2j}\gamma_{2j-1} \gamma_{2j}\right)\\
\Gamma'_{ext}&= 2^{-\frac{m-k +1}{2}} \prod_{j=k+1}^{m}\left( 1+i s_{2j-1,2j}\gamma'_{2j-1} \gamma'_{2j}\right)
\end{align}
The exterior pieces do not share any Majoranas, and so they commute. We can combine the interior pieces in such a way to almost remove them entirely. Note that, for projectors involving Majoranas
\begin{align}
&\left( 1+s_{2,3}i\gamma_2 \gamma_3\right)\left( 1+s_{1,2}i\gamma_1 \gamma_2\right)\nonumber\\
&\times\left( 1+s_{3,4}i\gamma_3 \gamma_4\right)\left( 1+i s_{2,3}i\gamma_2 \gamma_3\right)\nonumber\\
&=2\left( 1+s_{1,2}s_{2,3}s_{3,4}i\gamma_1 \gamma_4\right)\left( 1+s_{2,3}i\gamma_2 \gamma_3\right)
\end{align}
and we can use this as an induction step to remove all the projectors along the chain, 
\begin{align}
\Gamma_{int}\Gamma'_{int}P= 2^{-\frac{1}{2}}\left( 1+\left(\prod_{j=1}^{2k-1}s_{j,j+1}\right)i\gamma_{1} \gamma_{2k}\right)P \label{interior}
\end{align}
leaving only a coupling between the end points of the chain (figure \ref{intersection}).
\begin{figure}[htbp]
\begin{center}
\includegraphics[width=0.4\textwidth]{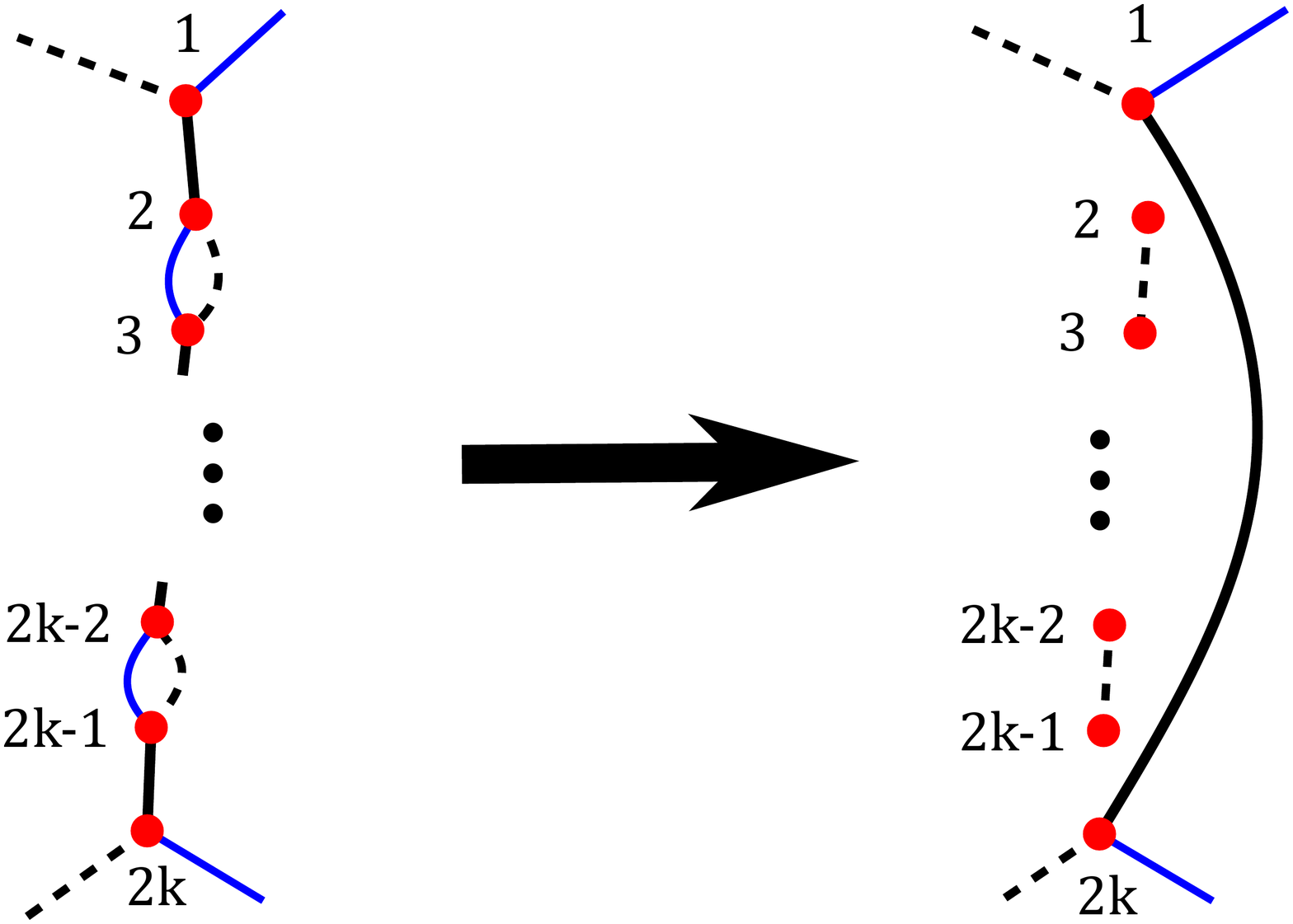}
\caption{Diagrammatic representation of the result in equation \ref{interior}. The projectors along the chain meld into a single projector enforcing $i\gamma_1\gamma_{2k} = \prod_{j=1}^{2k-1}s_{j,j+1}$}
\label{intersection}
\end{center}
\end{figure}

The remaining bilinear can also be absorbed using the following trick. We can pull bilinears of the form $s_{vw}i\gamma'_v\gamma'_w$ from both $P$ and $\Gamma'_{ext}$, both contain projectors which set those bilinears to 1. 
\begin{align}
\Gamma'_{ext}&i\gamma'_{1} \gamma'_{2k}P \nonumber\\
&=\Gamma'_{ext}s_{2k,2k+1}\gamma'_{1} \gamma'_{2k+1}P\nonumber\\
&=\Gamma'_{ext}s_{2k,2k+1}s_{2k+1,2k+2}i\gamma'_{1} \gamma'_{2k+2}P\nonumber\\
&\vdots \nonumber\\
&=\Gamma'_{ext}\left(\prod_{j=2k}^{2m-1}s_{j,j+1}\right)i\gamma'_{1} \gamma'_{2m}P\nonumber\\
&=\Gamma'_{ext}\left(-\prod_{j=2k}^{2m}s_{j,j+1}\right)P \label{exterior}
\end{align}
Combining equations \ref{interior} with \ref{exterior} allows us to conclude that
\begin{align}
\Gamma \Gamma' P &= 2^{-\frac{1}{2}} \left( 1-\prod_{j=1}^{2m}s_{j,j+1}\right)\Gamma_{ext} \Gamma'_{ext} P\nonumber\\
&=2^{\frac{1}{2}}\Gamma_{ext} \Gamma'_{ext} P
\end{align}
so long $\prod_{j=1}^{2m}s_{j,j+1} = -1$. This is, yet again, the Kasteleyn condition.

We see that we can absorb the action on the intersection into the surrounding projectors, so that the net result only depends on a chain of projectors along the exterior of the two plaquette terms. This is sufficient to show that the plaquette terms, as defined all the way back in equation \ref{eq:decomp}, commute exactly.
\end{appendix}

\bibliographystyle{unsrt}
\bibliography{Z2Refs}

\end{document}